\documentclass[10pt]{iopart}

\usepackage{graphicx}
\usepackage{hyperref} 
\usepackage{verbatim} 
\usepackage{epstopdf}
\usepackage[normalem]{ulem}
\usepackage{color}
\usepackage{dsfont}
\usepackage{float}
\usepackage{setspace}

\begin{document}
\title[Correlation properties of a three-body bosonic mixture in a harmonic trap]{Correlation properties of a three-body bosonic mixture in a harmonic trap}

\author{R E Barfknecht$ˆ{1,2}$, A S Dehkharghani$ˆ2$,
A Foerster$ˆ1$ and N T Zinner$ˆ2$}

\address{$ˆ1$ Instituto de F{\'i}sica, Universidade Federal do Rio Grande do Sul, Porto Alegre, Brazil}
\address{$ˆ2$ Department of Physics and Astronomy, Aarhus University, Aarhus, Denmark}

\begin{abstract}
\noindent We make use of a simple pair correlated wave function approach to obtain results for the ground state densities and momentum distribution of a one-dimensional three-body bosonic system with different interactions in a harmonic trap. For equal interactions this approach is able to reproduce the known analytical cases of zero and infinite repulsion. We show that our results for the correlations agree with the exact diagonalization in all interaction regimes and with analytical results for the strongly repulsive impurity. This method also enables us to access the more complicated cases of mixed interactions, and the probability densities of these systems are analyzed.
\end{abstract}
\submitto{\jpb}
\maketitle
\ioptwocol

\section{Introduction} Twenty years after the discovery of the phenomenon of the Bose-Einstein condensation \cite{BEC1D,BEC2D,BEC3D},the study of ultracold quantum gases continues to prosper with impact to various areas of physics. Meanwhile a large number of research branches have been established, revealing the beauty of the ultracold quantum world. One particular development has involved the optical confinement of ultracold atoms to one dimensional (1D) tubes, providing the realization of 1D quantum many body systems in the laboratory\cite{paredes,kinoshita,weiss,moritz,liao}. The emergence of successful experiments involving a small number of particles with high control and precision \cite{26D,27D,chinese,25Br} has generated an intense effort in the theoretical study of bosonic and fermionic few-body systems (see, for instance, \cite{31D,32D,33D,34D,brouzos,36D,37D,38D,39D,40D,41D}). The measurement of the effects of an impurity in an increasingly large fermionic system has also highlighted the discussion of the crossover between few and many-body systems \cite{brouzos_fewmany}. In this context, only recently experiments involving ultracold bosons with impurities have been performed \cite{26Z,27Z,28Z,29Z}, thus stimulating further theoretical investigations in the field. 

The presence of impurities in 1D bosonic environments generally turns its theoretical treatment more complicated, requiring innovative and skillful methods. The Bethe ansatz, for instance, which provides an exact solution for a system of bosons if all the interaction strengths are the same and the masses equal (Lieb-Liniger integrable model) is not applicable by introducing impurities, even in the homogeneous case \cite{35Br,36Br}.
So far not many techniques exist to handle few trapped bosons with impurities; we mention here the exact diagonalization \cite{bruno} and an analytical method in the strongly interacting limit \cite{zinner2}. Therefore, alternative approaches to explore and access the physics of such systems are highly welcome and constitute the main focus of the present article. 

Here we investigate the ground-state properties of a few body system of bosons in a one-dimensional harmonic trap in the presence of an impurity of the same mass but in a different hyperfine state. Experimentally, in these systems the interactions between the particles can be tuned via Feshbach resonances through the so-called confinement-induced resonances \cite{olshanii1}. We propose a trial wave function based on the analytical solution for two trapped particles \cite{busch}, which generalizes the pair correlated wave function approach \cite{brouzos} for the case of different interactions. This procedure allows us to study the few-body mixtures in a systematic way, and measurable quantities such as the correlation functions and momentum distributions can be determined. Through this method we obtain the momentum distribution and correlation functions for a system of two bosons and one impurity and show that our results for the correlations are in good agreement with the exact diagonalization for all interacting regimes and with existing analytical results for the strongly repulsive impurity limit. 

\section{Hamiltonian and Ansatz}

\subsection{Hamiltonian}
We start by considering a system of bosons with contact interactions in a one-dimensional harmonic trap. The Hamiltonian for the $N$-body case is written as
\begin{equation} 
\label{hamiltonian}
H=-\frac{1}{2}\sum_{i=1}^{N}\frac{\partial^2}{\partial x_{i}^2}+\sum_{i=1}^{N}\frac{x_i^2}{2}+\sum_{i<j} g_{ij}\delta(x_{i}-x_{j})
\end{equation}
where we consider the energies and lengths in units of $\hbar \omega_{L}$ and $b_{L}=\sqrt{\hbar/m\omega_{L}}$, with $\omega_{L}$ being the longitudinal harmonic confinement.
The parameter $g_{ij}$ is given in units of $\sqrt{m/\hbar^3\omega_{L}}$ and accounts for the possible different interactions between the pairs of atoms. Experimentally, these interactions can be controlled by means of a Feshbach resonance that gives rise to a so-called confinement-induced resonance. Since here we focus on neutral cold atoms, the interactions are typically of very short range, much shorter than the typical interparticle spacings in the system. Therefore, we can treat the atomic interactions as zero-range, even though they may be more complicated at very short distances. Experiments usually do not resolve the physics at the very small scales associated with atom-atom potential (as given
for instance by the van der Waals length). Furthermore, when the gas is confined into a 1D setup one may show that this also allows us to use a zero-range interaction (see for instance, \cite{olshanii1}). In the absence of a trapping potential and considering all interactions between the pairs to be equal, the Hamiltonian \ref{hamiltonian} reduces to the well known Lieb-Liniger integrable Hamiltonian \cite{lieb1,lieb2}, which exhibits a very rich physics and has been used to describe several experiments \cite{guan,batchelor}. In this work, we will be particularly interested in the case of three bosons with equal masses. \\

\subsection{Ansatz}
To calculate the physical properties of this model, we write our ansatz as 
\begin{equation}
\Psi(x_1,x_2,...,x_N)=\Phi_{CM}\psi_{R},
\end{equation}
where $\Phi_{CM}=\mathcal{N}_{CM}\exp{\left[\left(\sum_i^N x_i\right)^2/2N \right]}$ is the center of mass part of the wave function and $\mathcal{N}_{CM}$ is a normalization constant. The relative part of the wave function, $\psi_R$, is written as

\begin{equation}\label{eq2}
\psi_R=\mathcal{N}_R\prod_{i<j}^P D(\beta_{ij}|x_j-x_i|;\mu_{ij}),
\end{equation}
where $P=\frac{N(N-1)}{2}$ is the number of pairs, $D$ is a parabolic cylinder function \cite{gradshteyn} which depends on the absolute separation between the particles $|x_j-x_i|$ and the parameters $\beta_{ij}$ and $\mu_{ij}$, and $\mathcal{N}_R$ is a normalization constant. This separated form of the wave function has first been proposed and used to describe $N$-body bosonic systems with equal interactions in \cite{brouzos}, where it is employed to calculate energies and correlation properties. It is based on the seminal solution for the case of two bosons in a harmonic trap \cite{busch}. For the case of different interactions a similar approach has been proposed in \cite{39D,barfk} for the homogeneous case. Here we combine these two approaches to treat the general case of bosons with different interactions in a trap. Using the boundary condition for the delta function potential between a pair of particles, we find the following relation for $\mu_{ij}$ and $g_{ij}$:

\begin{equation}\label{eq3}
\frac{g_{ij}}{\beta_{ij}}=-\frac{2^{3/2}\Gamma(\frac{1-\mu_{ij}}{2})}{\Gamma(\frac{-\mu_{ij}}{2})},
\end{equation}
where $\Gamma$ are Gamma functions and $\mu_{ij}$ varies between 0 and 1 as $g_{ij}$ grows from 0 to $\infty$. By choosing all $\beta_{ij}=\sqrt{2/N}$ and considering all interactions equal ($g_{ij}=g$ for any pair) the total wave function reproduces the known analytical ground state results in the non-interacting 
\begin{equation}\label{eq4}
\Psi_0=\mathcal{N}e^{-\sum_ i^N \frac{x_i^2}{2}}
\end{equation}
and infinitely repulsive, also called Tonks-Girardeau (TG) \cite{TG}
\begin{equation}\label{eq5}
\Psi_{\infty}=\mathcal{N}e^{-\sum_ i^N \frac{x_i^2}{2}}\prod_{i<j}^P|x_j-x_i|
\end{equation}
cases, where $\mathcal{N}$ is the appropriate normalization constants for each limit. Notice that equation~\ref{eq5} holds only for the ground state of a system of identical bosons, mapping it into a spinless fermion system. It does not hold for strongly interacting fermions, where other approaches must be used \cite{fewbody}. Furthermore, for the particular case of $N=2$, equation (2) is also the exact wave function for any interaction strength \cite{busch}. Outside of these limits (e. g. for equal intermediate interactions or mixtures of bosons with different interactions) the parameters $\beta_{ij}$ are not fixed at $\sqrt{2/N}$, and we therefore treat them variationally.

\section{Probability Densities}
In this section we use our ansatz to calculate the densities for a system of three bosons in the harmonic trap. First we consider a system with equal interactions between the pairs in the limits of $g=0$ and $g\rightarrow \infty$, then we calculate the same properties for a system of two identical bosons plus an impurity (an atom that has a different interaction strength than the remaining pair). The case of four particles is briefly discussed in the end of this section.

\subsection{Equal Interactions}

The one-body correlation function is defined for a normalized wave function as 

\begin{equation}
\rho(x_{1})=\int dx_{2},...,dx_{N} |\Psi(x_{1},x_2,...,x_{N})|^2.
\end{equation}

\begin{figure}
\includegraphics[scale=0.6]{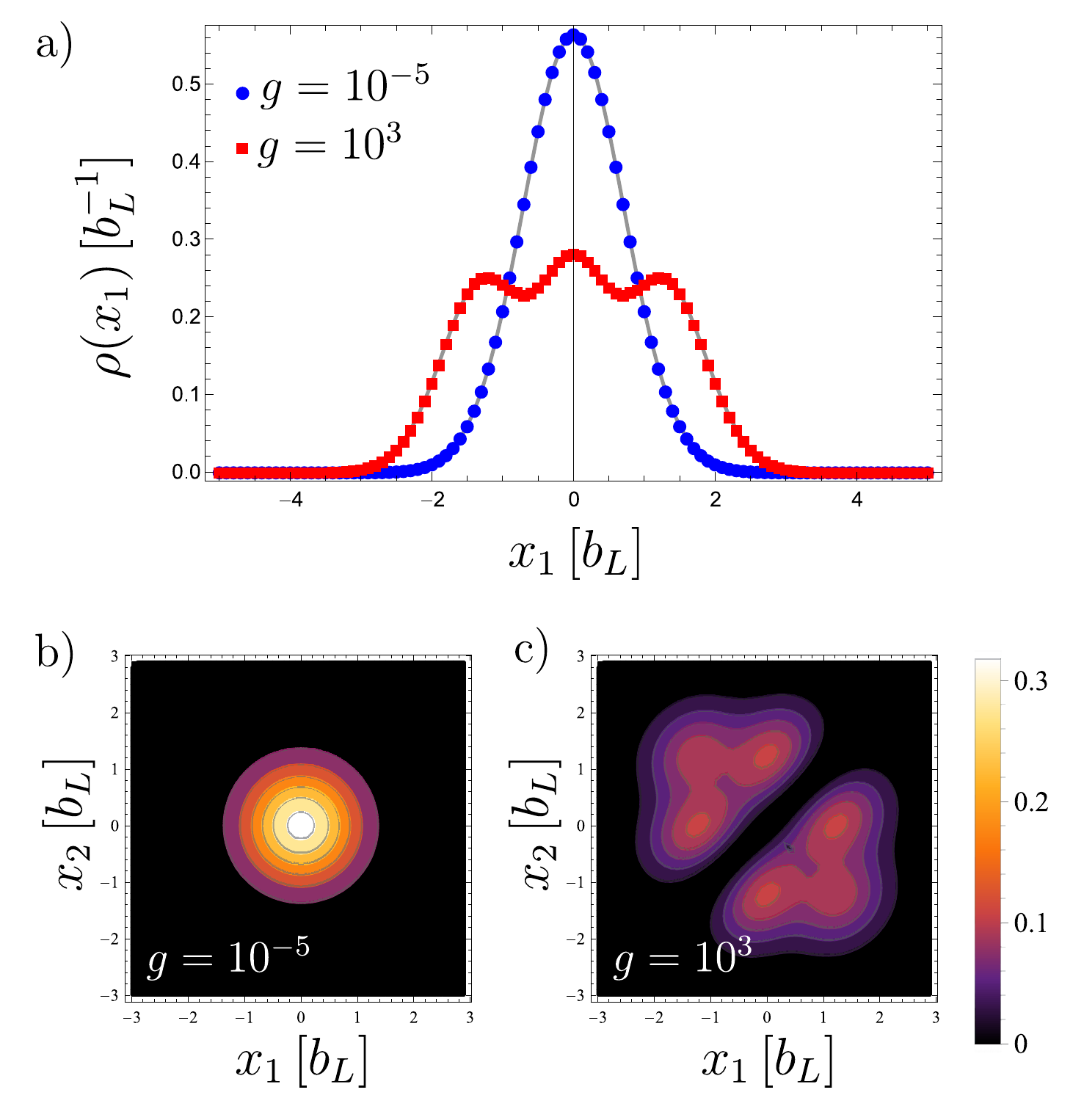} 
\caption{a) One-body correlation function in the non-interacting (blue circles) and strongly interacting (red squares) limits. The solid gray lines show the results obtained using the analytical limits \ref{eq4} and \ref{eq5}. Pair correlations for b) the non-interacting and c) the strongly repulsive limits.}
\label{fig1}
\end{figure}

In figure~\ref{fig1} a) we show results for this quantity obtained using $\Psi(x_1,x_2,...,x_N)=\Phi_{CM}\psi_R$ with $\psi_R$ defined in \ref{eq2}, for the strongly interacting ($g=1000$, $\mu \sim 1$) and for the non-interacting ($g=0$, $\mu=0$) limits in the case of three identical bosons, where our wave function reproduces the exact results. We observe the tendency of the atoms to separate in the trap in the strongly repulsive case, with the appearance of three separate peaks, one for each particle. The non-interacting case shows the expected Gaussian profile. We also notice that the wave function correctly reproduces the analytical cases given by \ref{eq4} and \ref{eq5}.
The pair correlation function, defined as $\rho_2(x_1,x_2)=\int dx_{3},...,dx_{N} |\Psi(x_{1},x_2,...,x_{N})|^2$, shows similar effects; in figure~\ref{fig1} b), the atoms do not interact; therefore the separation between any given pair can be zero. In figure.~\ref{fig1} c), the density goes to zero around the diagonal $x_{1}=x_{2}$, since the repulsion is strong. Results analogous to that of panel c) have been obtained in \cite{girardeau} for larger systems of identical particles.

\subsection{Different Interactions: Two Bosons and one Impurity}
\begin{center}
\begin{figure}[h]
\includegraphics[scale=0.5]{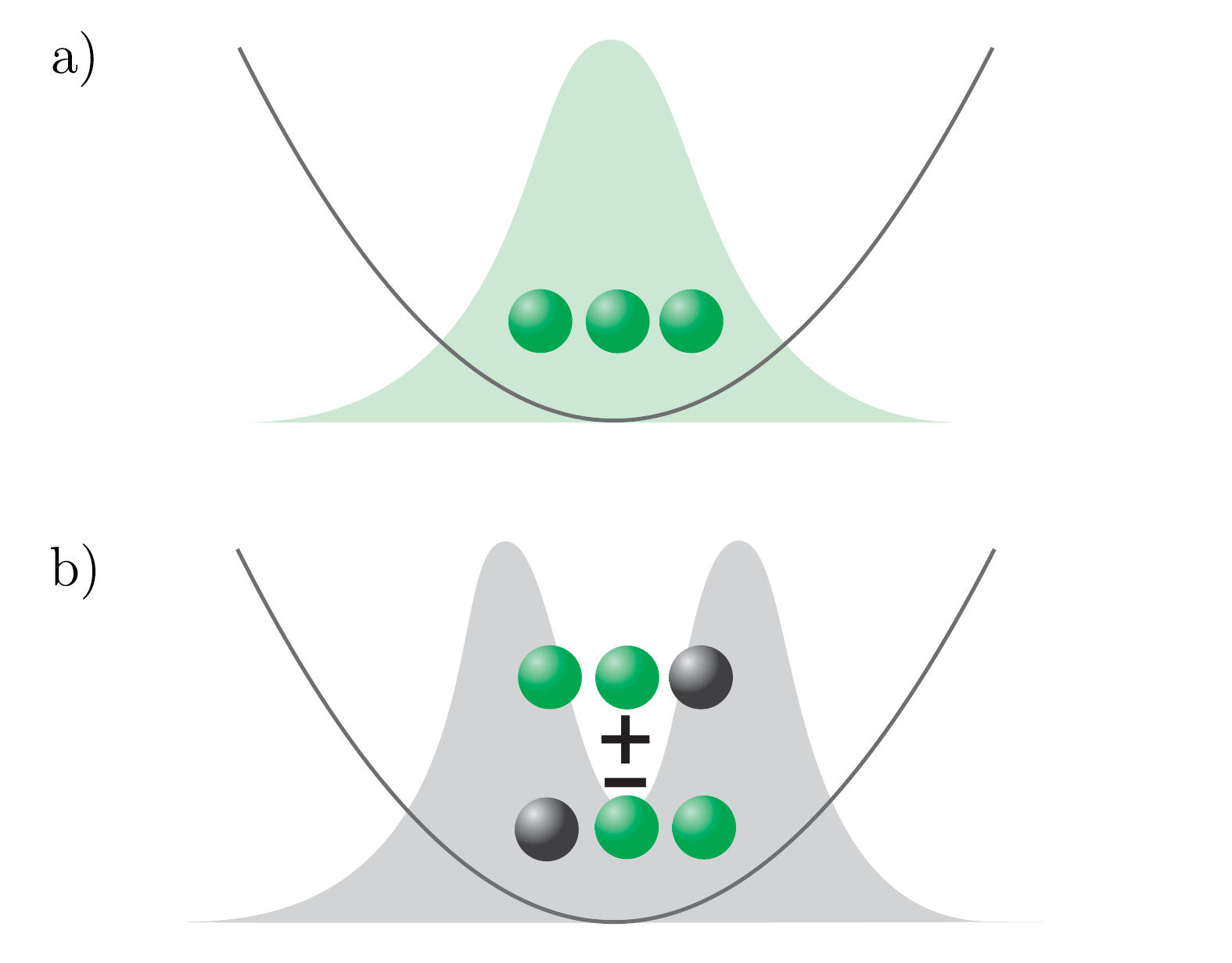} 
\caption{Depiction of two possible states for three bosons in the trap. a) Three identical, non-interacting bosons and b) two identical bosons and a strongly repulsive impurity.}
\label{fig2}
\end{figure}
\end{center}

We now turn to the case of different interactions between the atoms in the trap. We focus in the problem of two identical bosons plus an impurity atom. In figure~\ref{fig2} a) and b) we show a schematic depiction of two possible configurations of the system. In a) we have the non-interacting case where all bosons can be considered identical. In this scenario all particles tend to occupy positions close to the center of the trap. If the interaction between the impurity and the majority atoms is strong, as shown in figure~\ref{fig2} b), then the impurity will be found at the edges of the trap. This effect can be verified in the correlations of this system, as will be shown next.

We rewrite the coordinates as $x_I$ for the impurity atom and $(x_{M1},x_{M2})$ for the majority identical bosons. The interaction parameters are set as $g_{IM}$ for the impurity-majority interactions and $g_{MM}$ for the majority-majority interaction. The condition $\beta_{ij}=\sqrt{2/N}$ is no longer necessary in the case of different interactions. Furthermore, since the interaction $g$ may be different for each pair, the parameters $\beta_{ij}$ may also be varied independently. To improve the precision of our approach, we therefore treat $\beta_{ij}$ variationally, optimizing this parameter in each interaction case. We focus in three interaction regimes: non-interacting ($g\sim 0$), intermediate interaction ($g=2.56$) and strong interaction ($g=200$).

\begin{center}
\begin{figure}[h]
\includegraphics[scale=0.75]{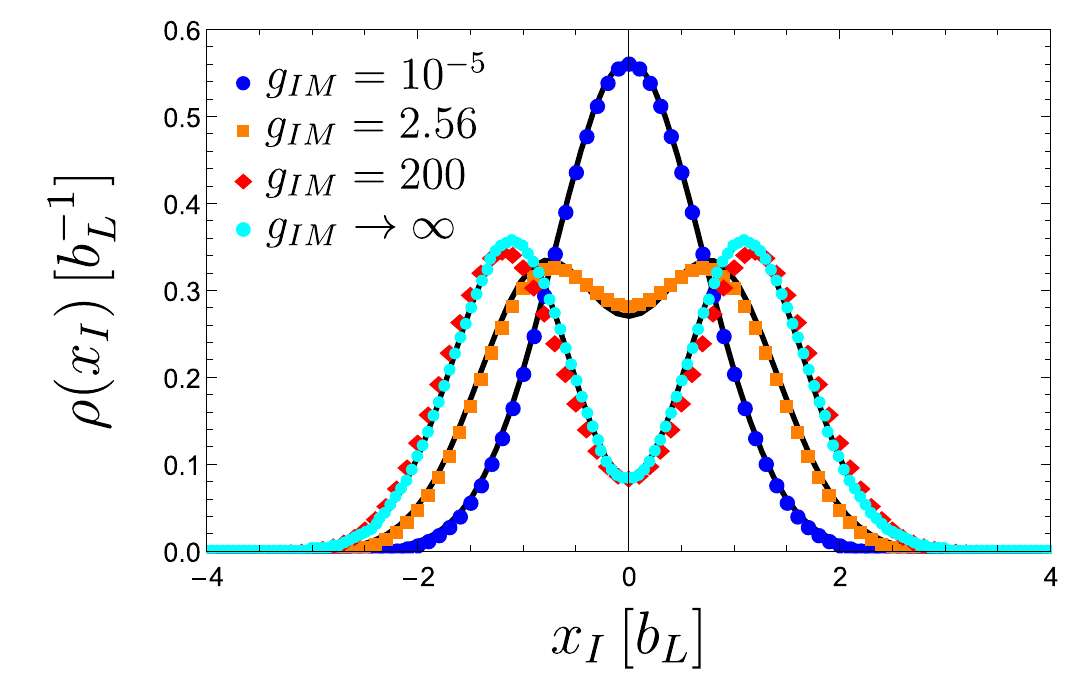} 
\caption{One-body correlation function for the impurity bosons with different impurity-majority interaction strengths. The black solid lines show the results obtained through exact diagonalization and the light blue dots the results from the analytical wave function in the infinite repulsive limit.}
\label{fig3}
\end{figure}
\end{center}

In figure~\ref{fig3} we plot the one-body correlation function for the impurity atom. We assume in this case that the majority bosons are non-interacting ($g_{MM}=0$). The most relevant physical effect in this case is the increasing separation of the density for the impurity, which tends to locate at one of the sides of the trap as a consequence of the repulsion with the majority bosons. This effect also shows the double degeneracy present in the ground state of systems such as these \cite{bruno,zinner1}. In figure~\ref{fig3} we also verify the validity of our approach by comparing to results obtained by exact diagonalization. In the non-interacting regime we observe once again that the exact density is reproduced. In the intermediate and strongly repulsive regimes, we observe small deviations from the exact results, although the general behavior is well captured. For the strongly repulsive regime we also present a comparison with an analytical wave function, obtained by considering $g_{MM}=0$ and $g_{IM}\rightarrow \infty$ (see \ref{app} for details). We notice that this result agrees well with the case of $g_{IM}=200$, both in the exact diagonalization and in the pair correlated ansatz, which shows that for a value of $g_{IM}=200$ the limit of infinite repulsion is already reached. The difference $\epsilon(x_{I})=(\rho(x_{I})_{ED}-\rho(x_{I}))^2$, where $\rho(x_{I})_{ED}$ is the results for the one-body correlation function obtained by exact diagonalization, is, as expected, $\epsilon(x_{I})=0$ for all points in the non-interacting case; in the interacting cases, it assumes slightly larger values, in particular around $x_{I}=0$ for $g_{IM}=2.56$ and $x_{I}\pm 1.8$ for $g_{IM}=200$. At these last points, nevertheless, the value of $\epsilon(x_{I})$ is still considerably small ($\sim 0.001$).

\begin{center}
\begin{figure}[h]
\includegraphics[scale=0.55]{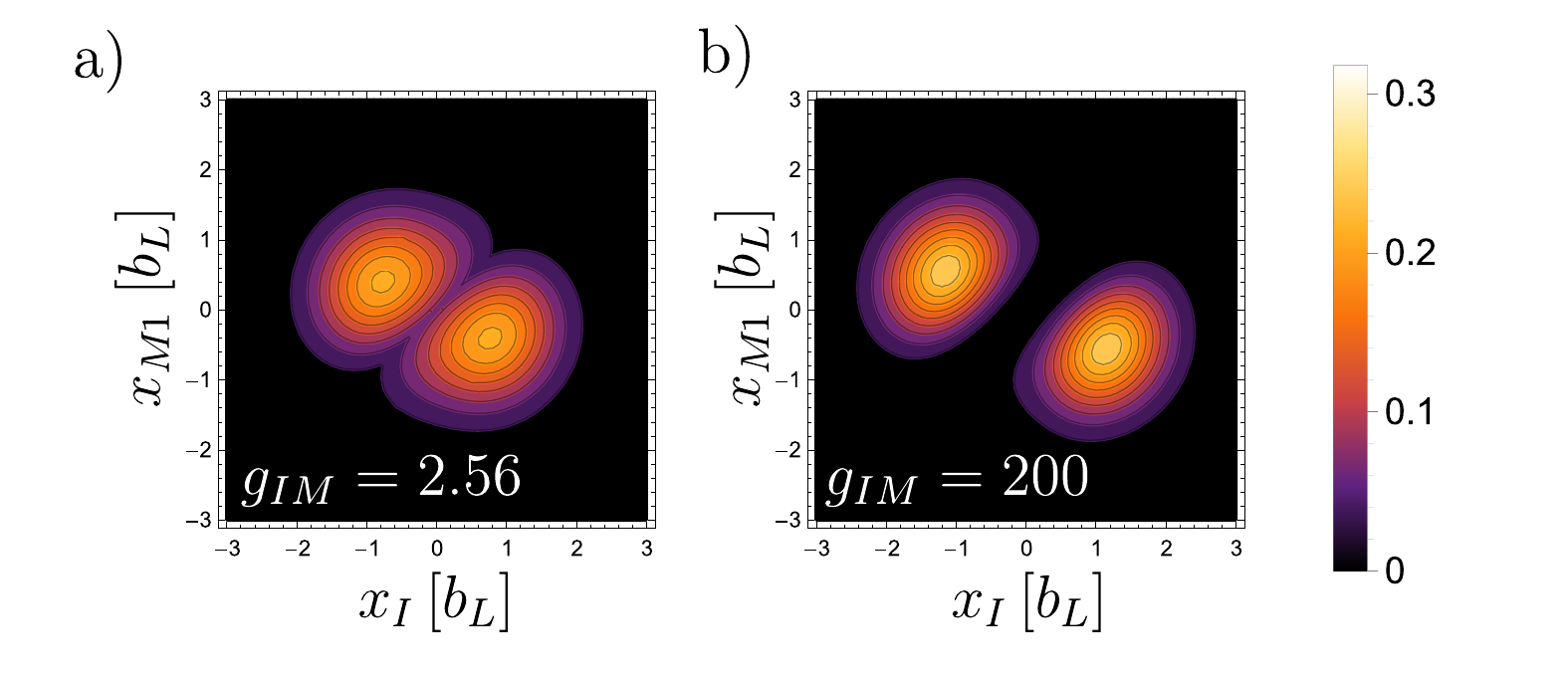} 
\caption{Pair correlation function for an impurity-majority pair, with interactions fixed as a) $g_{IM}=2.56$ and $g_{MM}=0$, b) $g_{IM}=200$ and $g_{MM}=0$.}
\label{fig4}
\end{figure}
\end{center}

\begin{center}
\begin{figure}
\includegraphics[scale=0.38]{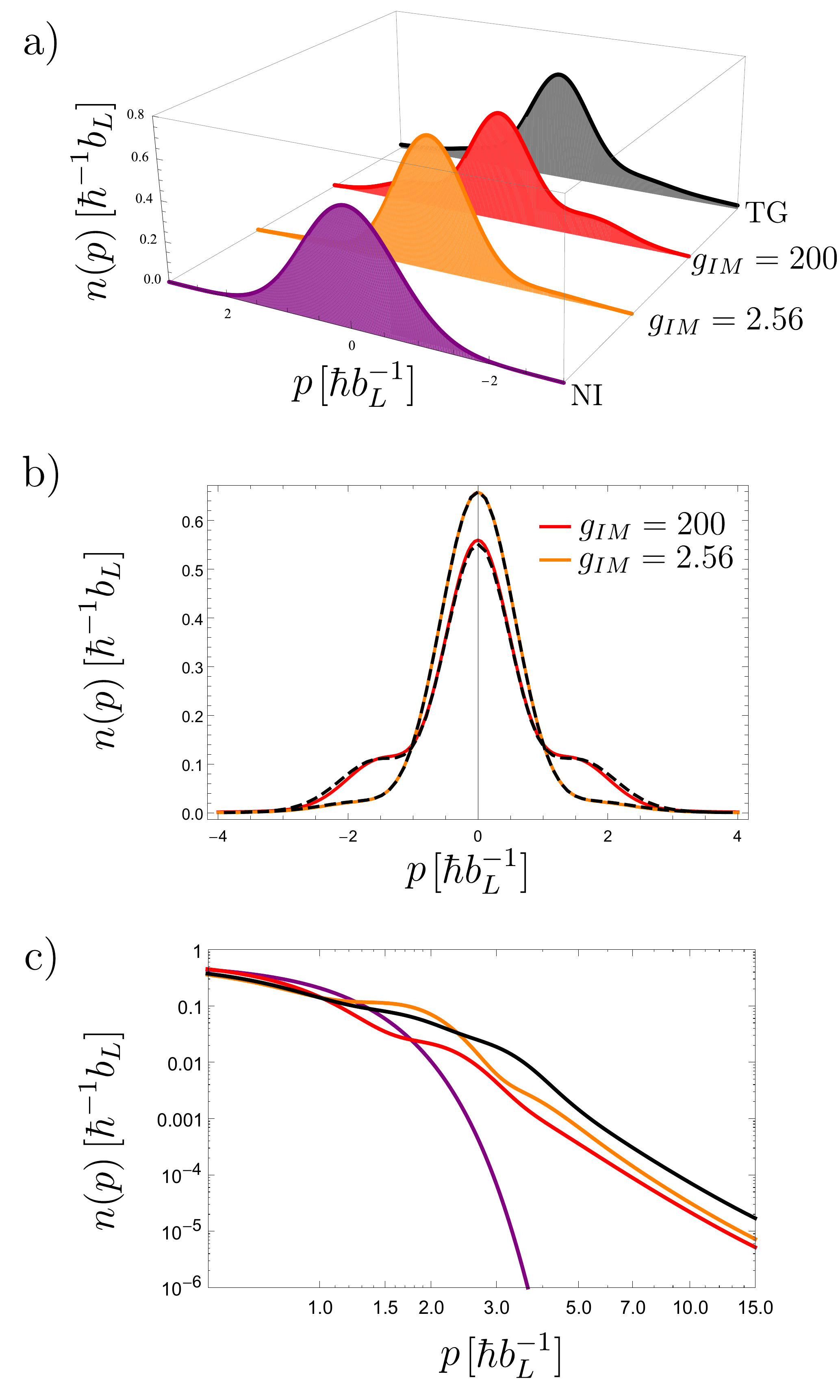} 
\caption{a) Dimensionless momentum distribution for the impurity in different interaction regimes. The purple curve shows the result obtained with the exact non-interacting wave function and the black curve shows the equal-interaction strongly repulsive case (Tonks-Girardeau gas). b) Same results as in a) for the mixed interactions: $g_{MM}=10^{-5}$ and $g_{IM}=200$ (red curve) and $g_{MM}=10^{-5}$ and $g_{IM}=2.56$ (orange curve), compared to exact diagonalization results (dashed curves). c) Log-log scale plot of the results in a). All interacting cases obey the $C/p^4$ power-law for asymptotic values of $p$.}
\label{fig5}
\end{figure}
\end{center}

The pair correlations shown for an impurity-majority pair in figures.~\ref{fig4} a) and b) also depict the tendency for the separation between the atoms of different species, along the diagonal $x_I=x_{M1}$, as the interaction $g_{IM}$ is increased. 
The momentum distribution, a quantity of great experimental interest, can be calculated as well from our approach, by taking a Fourier transform of the one-body correlation function: $n(p)=(1/2\pi)\int dx\,dx'\,e^{-ip(x-x')}\rho(x,x')$. In figure~\ref{fig5} a) we show results for the momentum distribution of the impurity as the interaction parameter $g_{IM}$ is increased. The non-interacting case (purple curve) shows a Gaussian profile that changes as the distribution gets larger around higher momentum values. This effect is more evident as the number of majority particles is increased \cite{zinner1}. 
In the mixed interactions cases we consider $g_{MM}=10^{-5}$. In figure~\ref{fig5} b) we show the agreement between these two results and the ones obtained by exact diagonalization.
In figure~\ref{fig5} b) we show the same results in log-log scale, where it becomes clear that in the interacting cases, the momentum distribution obeys a power law $C/p^4$ for high values of $p$. This is a characteristic behavior of systems with $\delta$ interaction; the constant $C$ is usually called the contact parameter, a concept that captures all universal properties of systems \cite{xiwen} and can be obtained analytically for both homogeneous \cite{olshanii2} and trapped \cite{minguzzi} models. Experimentally, however, the true nature of the two-body atomic interaction is expected to limit the applicability of this assumption for extremely large momenta (of the order of the inverse length scale of the two-body atomic interaction). The $1/p^4$ tail should be observable for large momenta that lie roughly between the inverse of the interparticle distance and the inverse van der Waals length.

\begin{figure}
\includegraphics[scale=0.42]{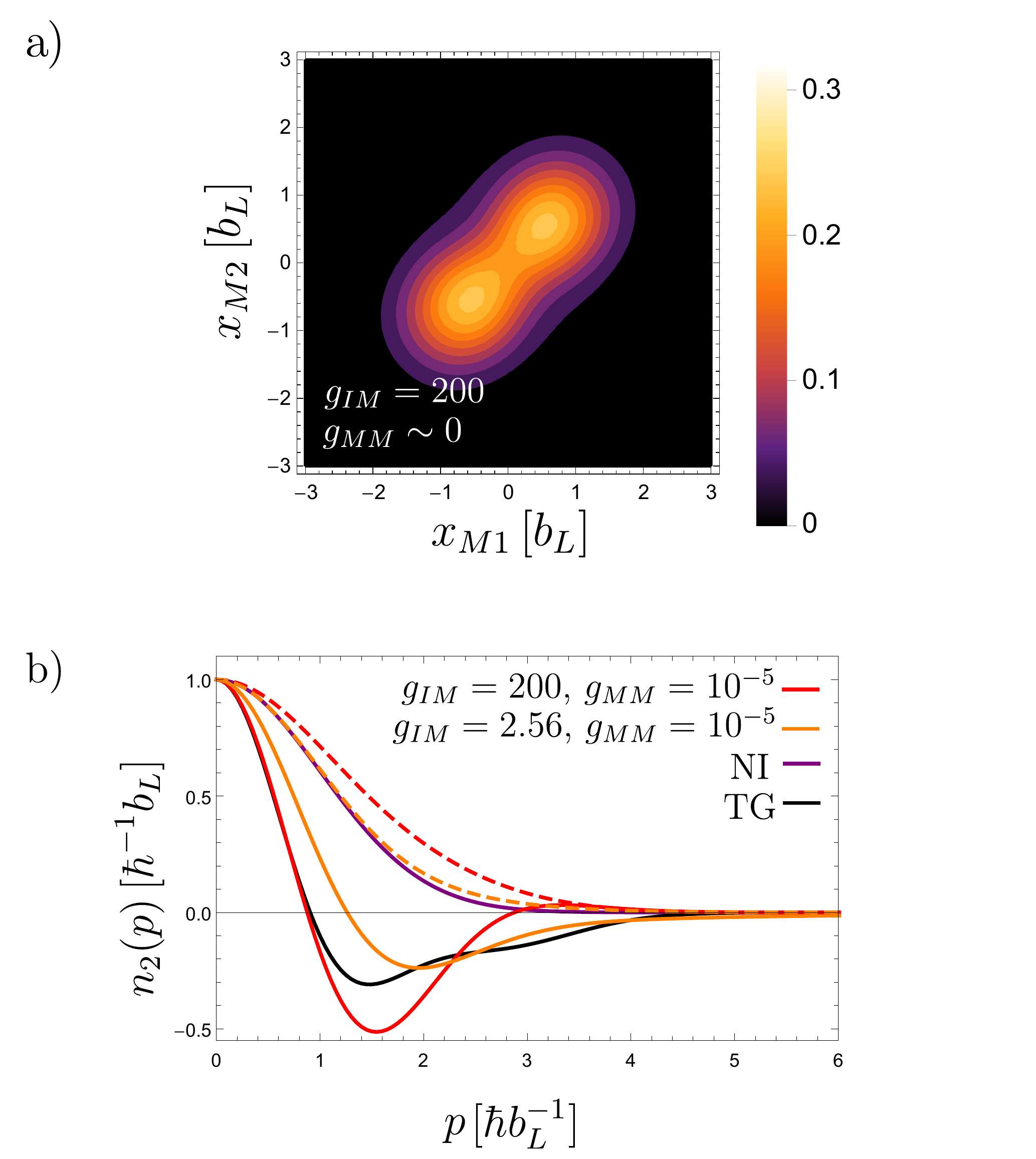} 
\caption{a) Correlations for the non-interacting majority-majority pair with strongly repulsive impurity. b) Fourier transform of the pair correlations in the non-interacting and Tonks-Girardeau cases (purple and black solid lines, respectively) and in the cases of intermediate (orange) and strong (red) repulsion by the impurity. Solid lines correspond to the impurity-majority pair, while the dashed lines correspond to the majority-majority pair.}
\label{fig6}
\end{figure}

For the particular case of a non-interacting majority pair and a strongly repulsive impurity it is also possible to notice the bunching of the majority bosons as a result of this repulsion. Figure~\ref{fig6} a) shows the pair correlations for the majority pair in this scenario. The identical bosons tend to occupy the same position due to the weak repulsion between them, but this position is slightly deviated from the origin of the system. This effect has consequences on the Fourier transform of the pair correlations, defined as $n_{2}(p)=\int dx_{3}\,e^{-ip(x_{1}-x_{2})}\rho_{2}(x_{1},x_{2})$ for an impurity-majority pair, and as $n_{2}(p)=\int dx_{1}\,e^{-ip(x_{2}-x_{3})}\rho_{2}(x_{2},x_{3})$ for a majority-majority pair. In figure~\ref{fig6} b) we show results for this quantity in the same interaction regimes as in figure~\ref{fig5}. The red and orange solid lines show the results for mixed interactions considering an impurity-majority pair, with a behavior similar to that of figure~\ref{fig5} when compared to the cases of equal interactions. The Fourier transform of the majority pair (dashed lines), however, assumes values larger than the non-interacting results for the low-momentum region. This effect can be traced back to peaks in the static structure factor of homogeneous systems \cite{barfk}, which accounts for an effective attractive interaction for the given pair of bosons.

\subsection{Three Bosons and one Impurity}

\begin{figure}[h]
\includegraphics[scale=0.725]{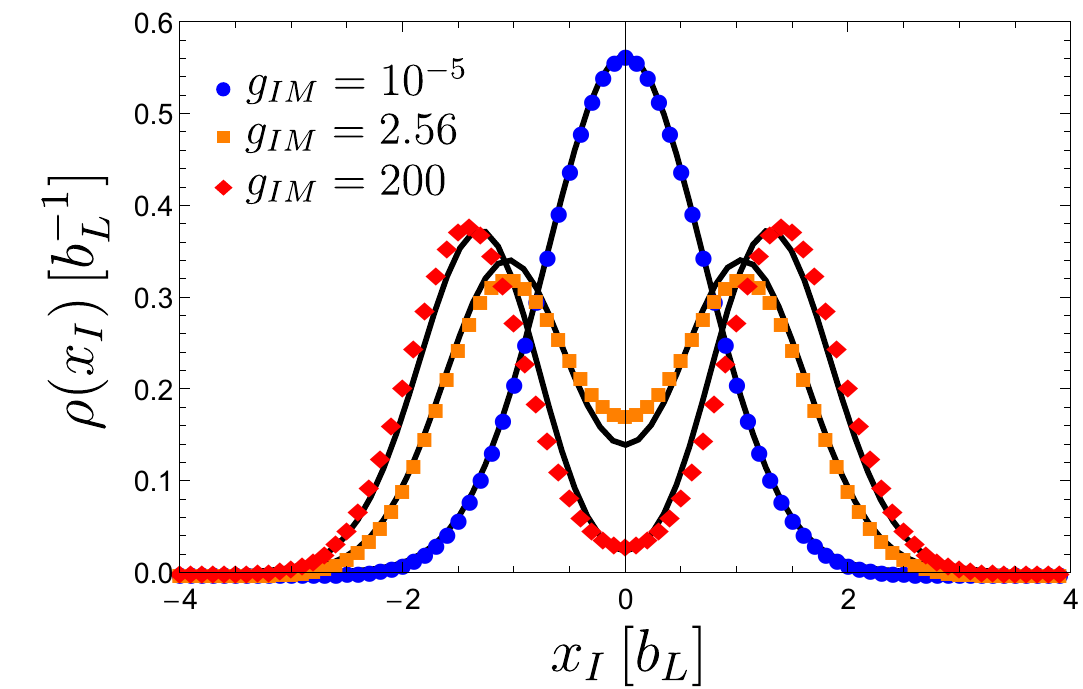} 
\caption{One-body correlation function for the impurity boson in the four particle system with three different impurity-majority interaction strengths. The black solid lines show the results obtained through exact diagonalization. In all cases we consider $g_{MM}=10^{-5}$.}
\label{fig7}
\end{figure}

Finally, to illustrate the generality of this approach, we extend it to the case of four particles (three identical bosons and an impurity). In figure~\ref{fig7}, we present results for the one-body densities, with interactions between the impurity and the majority pair again ranging from weak to strong, while the majority-majority interactions are kept small ($g_{MM}=10^{-5}$). Again, we notice the separation of the density for the impurity as the interactions are increased. The peaks are more pronounced than in figure~\ref{fig3}, since the number of identical bosons is larger. This effect can also be interpreted as precursor of ferromagnetism in bosonic systems \cite{zinner2}, since particles of the same species tend to bunch up on one side of the trap (provided that the intra-species interaction is small). 

\section{Conclusions}
We have used a pair correlated wave function, based on the solution for a pair of bosons in a trap, to access the features of the problem of few bosons in the presence of an impurity. We show that this wave function reproduces the limiting cases of zero interaction and infinite repulsion between the atoms. The results for intermediate interactions are consistent with the qualitative behavior expected for these systems. By increasing the interaction between the impurity and the remaining pair, we show that there is a tendency for this atom to occupy some position around the edges the trap. These effects on the one-body correlation functions of the system also reflect on quantities of experimental interest, such as the momentum distribution and the Fourier transform of the pair correlations. We also extend the approach to a case of three identical bosons and one impurity and compare results for the one-body densities with exact numerical results. It becomes clear that by combining a pair-correlated wave function with variational optimization it is possible to address other systems of great interest, such as balanced mixtures of two species of bosons or interacting ensembles beyond the ground state.

\ack A. Foerster and R. E. Barfknecht thank I. Brozous for inspiration and discussions. The folowing agencies - Conselho Nacional de Desenvolvimento Cient{\'i}fico e Tecnol{\'o}gico (CNPq), the Danish Council for Independent Research DFF Natural Sciences and the DFF Sapere Aude program - are gratefully acknowledged for financial support.

\appendix

\section{Deduction of the wave function for the infinitely repulsive impurity}\label{app}

In this appendix we present the wave function for a system of two non-interacting bosons and one infinitely repulsive impurity (see, for instance, \cite{zinner2}). This analytical wave function has been used to find the results shown in figure~\ref{fig3}. The Hamiltonian for a 2+1 bosonic systems can be written from \ref{hamiltonian} as
\begin{eqnarray}
H=&\frac{1}{2}(p_I^2+p_{M1}^2+p_{M2}^2)+ \frac{1}{2}(q_I^2+q_{M1}^2+q_{M2}^2) \nonumber \\ &+g_{IM}\delta(q_{I}-q_{M1})+g_{IM}\delta(q_{I}-q_{M2})\nonumber \\ &+ g_{MM} \delta(q_{M1}-q_{M2}),
\end{eqnarray}
where $q_I$ and $p_I$ denote the coordinate and momentum operators for the impurity, and $(q_{M1},q_{M2})$ and $(p_{M1},p_{M2})$ the coordinate and momentum operators for the majority bosons, respectively. Denoting $\mathbf{r}=(x,y,z)^T$ and $\mathbf{q}=(q_{M1},q_{M2},q_I)^T$, we can define the normalized Jacobi coordinates 
\begin{eqnarray}
\mathbf{r}=\mathbf{J}\mathbf{q}= \left[ \begin{array}{ccc}
\frac{1}{\sqrt{2}} & -\frac{1}{\sqrt{2}} & 0 \\
\frac{1}{\sqrt{6}} & \frac{1}{\sqrt{2}} & -\frac{\sqrt{2}}{\sqrt{3}} \\
\frac{1}{\sqrt{3}} & \frac{1}{\sqrt{3}} & \frac{1}{\sqrt{3}} \end{array} \right] 
\left[ \begin{array}{c}
q_{M1} \\
q_{M2} \\
q_{I} 
\end{array} \right],
\end{eqnarray}
with $\mathbf{J}\in SO(3)$. Notice that $z$ is the center-of-mass coordinate and $(x,y)$ are the relative coordinates. This transformation allows us to rewrite the hamiltonian in terms of the new variables: 
\begin{eqnarray}
H=H_0+V,
\end{eqnarray}
where 
\begin{equation}
H_0=\frac{1}{2}(\mathbf{k}^2+\mathbf{r}^2),
\end{equation}
with $\mathbf{k}=\mathbf{J}\mathbf{p}$ transformed correspondingly and
\begin{eqnarray}
V&=\frac{1}{\sqrt{2}}\left[ g_{MM}\delta(x) +g_{IM}\delta\left(-\frac{1}{2}x+\frac{\sqrt{3}}{2}y\right)\right. \nonumber \\
&+ \left. g_{IM} \delta\left(-\frac{1}{2}x-\frac{\sqrt{3}}{2}y \right)\right].
\end{eqnarray}
This hamiltonian has solutions that are separable between the center-of-mass and relative motion. The center-of-mass solution is given simply by the one-dimensional harmonic oscillator wave functions:
\begin{equation}
\psi_{\eta}(z)=\frac{1}{\sqrt{2^\eta \eta!}}\left(\frac{1}{\pi}\right)^{1/4}e^{-z^2/2}H_\eta (z),
\end{equation}
where $H_\eta (z)$ are the Hermite polynomials. For the relative motion we can also define hyperspherical coordinates as
\begin{eqnarray}
\rho=\sqrt{x^2+y^2}\nonumber \\
\tan(\phi)=y/x,
\end{eqnarray}
where $\rho \in [0,\infty) $ and $\phi \in [0,2\pi]$. The delta functions can be written as
\begin{eqnarray}
\delta\left(-\frac{1}{2}x+\frac{\sqrt{3}}{2}y\right)=\frac{1}{\rho}\left(\delta\left(\phi-\frac{\pi}{6}\right)+\delta\left(\phi-\frac{7\pi}{6}\right)\right) \nonumber \\
\end{eqnarray}
and the center-of-mass separated hamiltonian in polar coordinates becomes:
\begin{eqnarray}
H_0=\frac{1}{2}\left(-\frac{1}{\rho}\frac{\partial}{\partial \rho}-\frac{\partial^2}{\partial \rho^2}-\frac{1}{\rho^2}\frac{\partial}{\partial \phi^2}+\rho^2\right).
\end{eqnarray}
The system can now be divided into $N!=6$ regions concerning the possible particle orderings, namely:
\begin{table}[H]
\centering
\label{my-label}
\begin{tabular}{ll}
\textbf{I}: & $q_{M1}>q_{M2}>q_I$ \\
\textbf{II}: & $q_{M1}>q_{I}>q_{M2}$ \\
\textbf{III}: & $q_{I}>q_{M1}>q_{M2}$ \\
\textbf{IV}: & $q_{I}>q_{M2}>q_{M1}$ \\
\textbf{V}: & $q_{M2}>q_{I}>q_{M1}$ \\
\textbf{VI}: & $q_{M2}>q_{M1}>q_I$
\end{tabular}
\end{table}
\noindent
The solutions for the relative part of the wave function are given by
\begin{eqnarray}
\psi_{\nu,\mu}(\rho,\phi)=\mathcal{N}\,U(-\nu,\mu+1,\rho^2)e^{-\rho^2/2}\rho^\mu f_\mu(\phi),
\end{eqnarray} 
where $\nu,\mu=0,1,2...$, $U(-\nu,\mu+1,\rho^2)$ are the confluent hypergeometric functions \cite{abramowitz} and $\mathcal{N}$ is a normalization factor. We make an ansatz for the function $f_\mu(\phi)$ in regions \textbf{I}
\begin{equation}
f_\mu(\phi)=c\cos(\mu\phi)+d\sin(\mu\phi),
\end{equation}
and \textbf{II}
\begin{equation}
f\mu(\phi)=a\cos(\mu\phi)+b\sin(\mu\phi).
\end{equation}
By symmetry considerations, it is enough to determine the coefficients at these two regions. For a given choice of $\rho=\rho_0$, at the contact points between the particles the delta potential introduces a derivative discontinuity of the kind
\begin{equation}
\Delta\left(\frac{\partial f(\mu,\phi)}{\partial \phi}\right)=G_{IM}f(\mu,\phi),
\end{equation}
where 
\begin{equation}
\Delta\left(\frac{\partial f(\mu,\phi)}{\partial \phi}\right)\equiv \left(\frac{\partial \psi}{\partial \phi}\bigg|_+ -\frac{\partial \psi}{\partial \phi}\bigg|_-\right),
\end{equation}
and $G_{IM}\equiv \sqrt{2}g_{IM}\rho_0$. This last parameter becomes independent of $\rho_0$ as $g_{IM}\rightarrow \infty$. By exploiting these conditions, the continuity of the wave function at the six contact points, and choosing $g_{MM}=0$, we arrive finally at the wave function for the two lowest states in the case of infinitely repulsive impurity and two non-interacting bosons:
\begin{eqnarray}
\psi(z,\rho,\phi)&=Ce^{-z^2/2}\rho^{3/2}e^{-\rho^2/2}\times \nonumber\\&\left\{
\begin{array}{lr}
\sin(\frac{3}{2}(\phi+\frac{\pi}{6})) &: \phi \in \textbf{III}\\
0 &: \phi \in \textbf{II}\\
\mp\sin(\frac{3}{2}(\phi-\frac{\pi}{6})) &: \phi \in \textbf{I}\\
\end{array}
\right.
\end{eqnarray} 
(-) for the ground state with even parity, and (+) for the 1st excited state with odd parity. Notice that the wave function is symmetric in the regions \textbf{I} and \textbf{IV}, \textbf{II} and \textbf{V} and finally \textbf{III} and \textbf{VI}. From here one can easily reproduce the result obtained in figure~\ref{fig3} in the main text.

\section{Total density for mixed interactions}\label{app2}

\begin{figure*}
\includegraphics[scale=0.7]{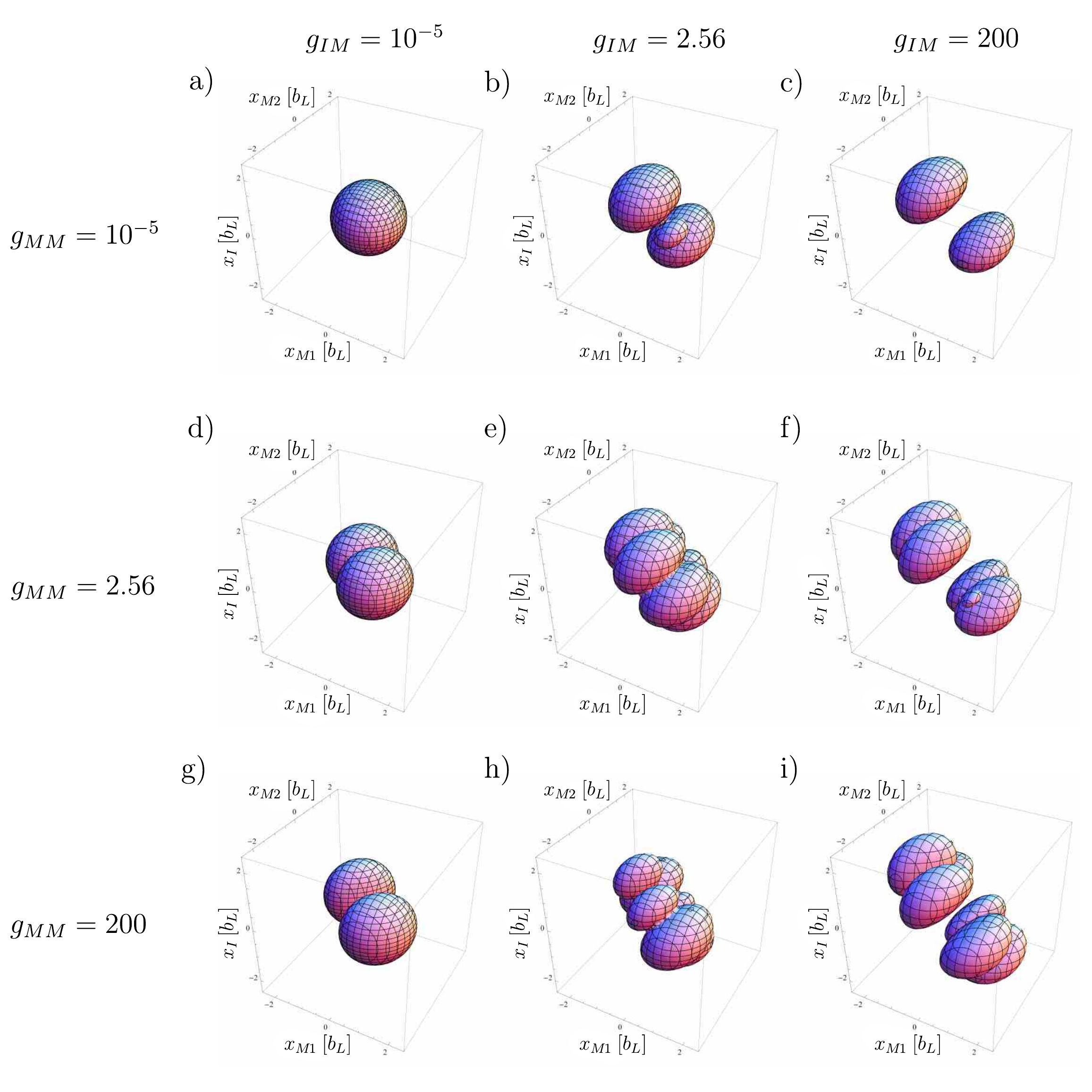} 
\caption{Normalized total density $|\Psi(x_I,x_{M1},x_{M2})|^2$ shows the depletion of the probability along the contact manifolds for different interactions. The repulsion between the majority-majority pair and the impurity-majority pairs is increased from top to bottom and left to right, respectively.}
\label{fig8}
\end{figure*}

Considering our approach has been validated by the comparison to exact numerical results, and to further elucidate the effect of the repulsive delta interactions between the pairs on the correlations, we now look at the normalized total density $|\Psi(x_I,x_{M1},x_{M2})|^2$ in nine different situations. 
The panels a), e) and i), along the diagonal in figure~\ref{fig8} have equal interactions between all the pairs. For the non-interacting a) and strongly repulsive i) densities we approximately reproduce the two cases shown in figure~\ref{fig1} a). Panel e) shows the intermediate case, with the depletion of probability along the manifolds $\{x_{I}=x_{M1}$, $x_{I}=x_{M2}$, $x_{M1}=x_{M2}\}$, in a similar way as shown in \cite{brouzos}. The off-diagonal densities are clearly not symmetric: in the column a), d) and g), only the interaction $g_{MM}$ is being increased. That means the probability is lowered only along the manifold $\{x_{M1}=x_{M2}\}$, which leads to the separation of the density in two lobes. On the other hand, the line a), b), c) shows depletion of the probability on two manifolds, namely $\{x_{I}=x_{M1}$, $x_{I}=x_{M2}\}$, which leads to the appearance of a smaller lobe on b). The mixed cases f) and h) also reflect the asymmetry of the interactions on the probability densities.

\end{document}